\documentclass{article}

\usepackage{arxiv}

\usepackage[utf8]{inputenc} % allow utf-8 input
\usepackage[T1]{fontenc}    % use 8-bit T1 fonts
\usepackage{lmodern}        % https://github.com/rstudio/rticles/issues/343
\usepackage{hyperref}       % hyperlinks
\usepackage{url}            % simple URL typesetting
\usepackage{booktabs}       % professional-quality tables
\usepackage{amsfonts}       % blackboard math symbols
\usepackage{nicefrac}       % compact symbols for 1/2, etc.
\usepackage{microtype}      % microtypography
\usepackage{graphicx}

\title{Current state and prospects of R-packages for the design of
experiments}

\author{
    Emi Tanaka
    \thanks{Corresponding author}
   \\
    Department of Econometrics and Business Statistics \\
    Monash University \\
  Clayton, VIC 3800 \\
  \texttt{\href{mailto:emi.tanaka@monash.edu}{\nolinkurl{emi.tanaka@monash.edu}}} \\
   \And
    Dewi Amaliah
   \\
    Department of Econometrics and Business Statistics \\
    Monash University \\
  Clayton, VIC 3800 \\
  \texttt{} \\
  }

\providecommand{\tightlist}{%
  \setlength{\itemsep}{0pt}\setlength{\parskip}{0pt}}

\newlength{\cslhangindent}
\newlength{\csllabelwidth}
\newlength{\cslentryspacingunit} % times entry-spacing
  {}%
  {\par}
\newenvironment{CSLReferences}[2] % #1 hanging-ident, #2 entry spacing
 {% don't indent paragraphs
  \setlength{\parindent}{0pt}
  \ifodd #1
  \let\oldpar\par
  \def\par{\hangindent=\cslhangindent\oldpar}
  \fi
  \setlength{\parskip}{#2\cslentryspacingunit}
 }%
 {}
\usepackage{calc}

\usepackage{colortbl}
\usepackage{bera}
\usepackage{xcolor}
\usepackage{hyperref}
\usepackage[utf8]{inputenc}
\def\tightlist{}

\begin{document}
\maketitle

\begin{abstract}
Re-running an experiment is generally costly and, in some cases,
impossible due to limited resources; therefore, the design of an
experiment plays a critical role in increasing the quality of
experimental data. In this paper, we describe the current state of
R-packages for the design of experiments through an exploratory data
analysis of package downloads, package metadata, and a comparison of
characteristics with other topics. We observed that experimental designs
in practice appear to be sufficiently manufactured by a small number of
packages, and the development of experimental designs often occurs in
silos. We also discuss the interface designs of widely utilized R
packages in the field of experimental design and discuss their future
prospects for advancing the field in practice.
\end{abstract}

\hypertarget{introduction}{%
\section{Introduction}\label{introduction}}

The critical role of data collection is well captured in the expression
``garbage in, garbage out'' -- in other words, if the collected data are
rubbish, then no analysis, however complex it may be, can make something
out of it. Therefore, a carefully crafted data-collection scheme is
critical for optimizing the information from the data. The field of
experimental design is specifically devoted to planning the collection
of experimental data, largely based on the founding principles of Fisher
(1935) or an optimization framework like those described in Pukelsheim
(2006). These experimental designs are often constructed with the aid of
statistical software such as R (R Core Team 2021), Python (Rossum 1995),
and SAS (SAS Institute 1985); thus the use of experimental design
software can inform us about some aspects of experimental designs in
practice.

Methods for data collection can be dichotomized by the type of data
collected -- namely, experimental or observational -- or alternatively,
categorized as experimental design (including quasi-experimental design)
or survey design. This dichotomization, to a great extent, is seen in
the \href{https://cran.r-project.org/web/views/}{Comprehensive R Archive
Network (CRAN) task views} (a volunteer maintained list of R-packages by
topic) where R-packages for experimental design are in
\href{http://CRAN.R-project.org/view=ExperimentalDesign}{\emph{ExperimentalDesign}}
task view and R-packages for survey designs are in
\href{http://CRAN.R-project.org/view=OfficialStatistics}{\emph{OfficialStatistics}}
task view. A full list of available topics is provided in Table S1 in
the Supplementary Materials. A subset of experimental designs is
segregated into the
\href{http://CRAN.R-project.org/view=ClinicalTrials}{\emph{ClinicalTrials}}
task view, where the focus is on clinical trials with primary interest
in sample size calculations. This paper focuses on packages in
\href{http://CRAN.R-project.org/view=ExperimentalDesign}{\emph{ExperimentalDesign}}
task view, henceforth referred to as ``DoE packages''.

From the
\href{http://CRAN.R-project.org/view=ExperimentalDesign}{\emph{ExperimentalDesign}}
task view, there are 105 R packages for the experimental design and
analysis of data from experiments. The sheer quantity and variation of
experimental designs in the R-packages are arguably unmatched with any
other programming languages, for example, in Python, only a handful of
packages that generate design of experiment exist (namely
\texttt{pyDOE}, \texttt{pyDOE2}, \texttt{dexpy}, \texttt{experimenter},
and \texttt{GPdoemd}) with a limited type of design. Thus, the study of
DoE packages, based on quantitative and qualitative data, can provide an
objective view of the state of current experimental designs in practice.

The utility of the software can also be described by its design to
facilitate the clear expression and interpretation of the desired
experimental design. Certain programming language designs can hinder or
discourage the development of reliable programs (Wasserman 1975). The
immense popularity of
\href{https://cran.r-project.org/web/packages/tidyverse/index.html}{\texttt{tidyverse}}
(a collection of R-packages for various stages of data analysis that
places enormous emphasis on the interface design by Wickham et al. 2019)
is a testament to the impact that an interface design can have in
practice. The practice of experimental design can be advanced by
adopting similar interface design principles across the DoE packages.

The remainder of this paper is organized as follows. Section \ref{data}
briefly describes the data source used for the analysis; Section
\ref{eda} presents some insights into the state of the current DoE
packages by the exploratory data analysis of package download data, text
descriptions and comparisons with other CRAN task views; Section
\ref{design} discusses the interface designs of widely used DoE
packages, and we conclude with a discussion in Section \ref{discussion}
of future prospects in the software development of experimental designs.

\hypertarget{data}{%
\section{Data}\label{data}}

To study the DoE packages, we analyse data using three sources of data
as described below.

\hypertarget{rstudio-cran-download-logs}{%
\subsection{RStudio CRAN download
logs}\label{rstudio-cran-download-logs}}

The Comprehensive R Archive Network (CRAN) is a network of servers
located across the world that stores mirrored versions of the R and R
packages. The most popular network is the RStudio mirror (the default
server for those that use the RStudio IDE). The RStudio mirror is also
the only server that provides a comprehensive daily download logs of R
and R packages since October 2012. The summary data can be easily
accessed using the
\href{https://cran.r-project.org/web/packages/cranlogs/index.html}{\texttt{cranlogs}}
package (Csárdi 2019). This paper uses the data from the beginning of
2013 to the end of 2021 (a total of nine years) for the packages in the
CRAN task views.

\hypertarget{package-descriptions}{%
\subsection{Package descriptions}\label{package-descriptions}}

All CRAN packages have a title, description, package connections
(suggests, depends, and imports of other packages), and other
meta-information in the DESCRIPTION file. We use text data from the
title and description (accessed in 2022-12-12).

\hypertarget{cran-task-views}{%
\subsection{CRAN task views}\label{cran-task-views}}

CRAN task views are volunteer-maintained lists of R-packages on CRAN
relevant to the corresponding topic. There were 39 CRAN task views in
total. Table S1 in the Supplementary Materials list the available topics
from the
\href{https://cran.r-project.org/web/packages/ctv/index.html}{\texttt{ctv}}
package (Zeileis 2005). The list of packages in each CRAN task view (as
of 2022-12-12) is used to contrast the characteristics of the DoE
packages.

\hypertarget{eda}{%
\section{Exploratory data analysis}\label{eda}}

In this section, we derive some conjectures based on an analysis of the
data described in Section \ref{data}. All results presented are from
exploratory data analysis of observational data, consequently, all
interpretations are somewhat speculative and may not be indicative of
the true state of the field of experimental design. In particular, any
analysis over time is confounded by the fact that the nature of users
and package management has changed over the years. It should be noted
that some DoE packages may have been archived or removed from the task
view over the years; therefore, any cross-sectional analysis presented
may not reflect the set of all DoE packages at that particular time
period (although we assume such incidences are low).

A subset of DoE packages is not primarily about the design of
experiments but about the analysis of experimental data. A complete
delineation of these packages is difficult, as there is almost always at
least one function that can aid decisions or constructions of
experimental designs (and any categorization is prone to our subjective
bias); therefore we opted not to remove any DoE packages in the
analysis.

\hypertarget{small-but-diverse-set-of-packages-are-sufficient-for-most-experimental-designs-in-practice}{%
\subsection{Small, but diverse, set of packages are sufficient for most
experimental designs in
practice}\label{small-but-diverse-set-of-packages-are-sufficient-for-most-experimental-designs-in-practice}}

\begin{figure}[htbp]

{\centering \includegraphics{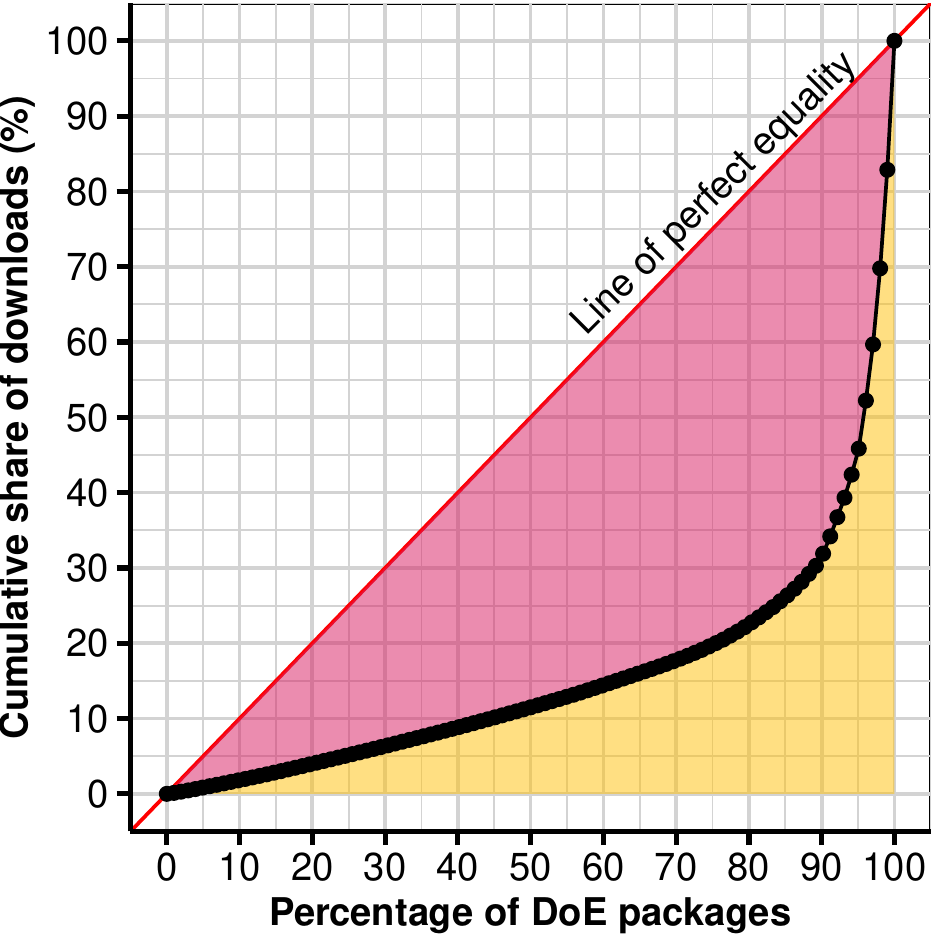} 

}

\caption{Lorenz curve of the total download count for DoE packages in 2021. The red line corresponds to the line of perfect equality. The yellow region shows the area under the Lorenz curve and the red region shows the area of the gap in equality.}\label{fig:plot-lorenz}
\end{figure}

There are at least 50 DoE packages since 2013 but most of the downloads
are concentrated in only a handful of packages. For example, Figure
\ref{fig:plot-lorenz} shows a Lorenz curve (Lorenz 1905) for the total
package downloads in 2021 for 102 DoE packages (first released prior to
2021). We can see from Figure \ref{fig:plot-lorenz} that the bottom 90\%
of DoE packages (in terms of total download count in 2021) only share
approximately 32\% of total downloads across all DoE packages; in other
words, 68\% of the total downloads are due to 10 packages (10\% of the
DoE packages).

\begin{figure}[htbp]

{\centering \includegraphics{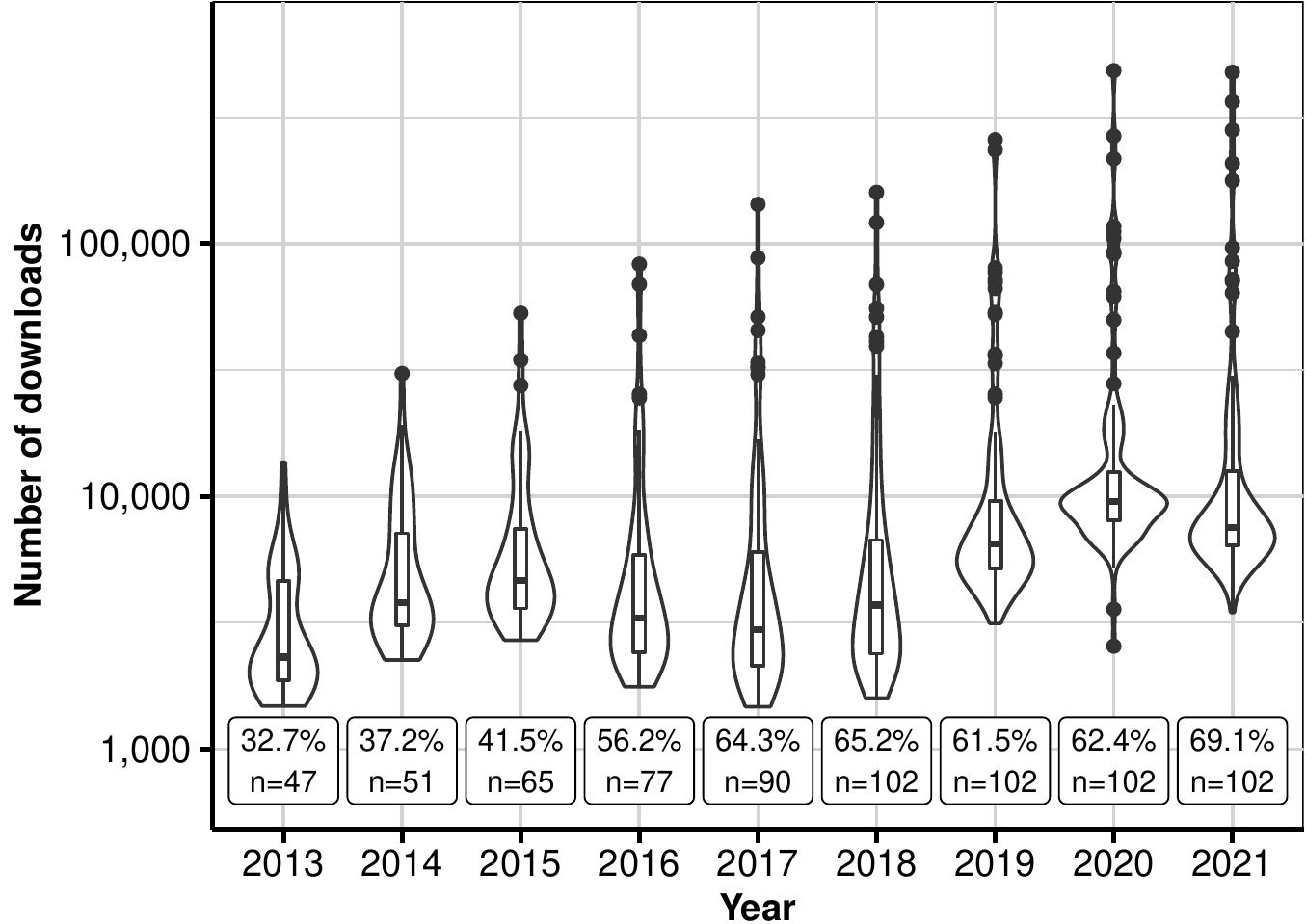} 

}

\caption{Distribution of number of downloads for DoE packages by year. Packages were removed in any year if they were released in that year or later so that each download count was for the full year. The label at the bottom of the plot shows the Gini index for downloads and the number of packages with a full download count in the corresponding year. In the last six years, the Gini index has consistently exceeded 60\%, indicating that most downloads are due to a relatively small number of packages. }\label{fig:download-share}
\end{figure}

If we consider package downloads as a measure of ``wealth'', then we can
consider using the Gini index (Gini 1921) as a measure of download
inequality across packages. The ratio of the red region to the total
colored regions in Figure \ref{fig:plot-lorenz} corresponds to the Gini
index for 2021. A Gini index of 0\% indicates equality in downloads
across packages whereas a value of 100\% indicates maximal inequality
(all downloads are due to one package). In Figure
\ref{fig:download-share}, we see that the distributions of the package
downloads each year have a heavy right tail with the Gini index ranging
from 32.7\% to 69.1\% across the years 2013 to 2021, indicating that
there is a high level of inequality in package downloads, particularly
with more pronounced inequality in the last six years.

An increase in the number of packages that are not highly downloaded may
mean that \textbf{\emph{there are more packages to construct niche
experimental designs}}. Some examples of these packages include
\href{https://cran.r-project.org/web/packages/qtlDesign/index.html}{\texttt{qtlDesign}},
\href{https://cran.r-project.org/web/packages/PwrGSD/index.html}{\texttt{PwrGSD}}
and
\href{https://cran.r-project.org/web/packages/Crossover/index.html}{\texttt{Crossover}}
made for QTL experiments, group sequential designs and crossover trials,
respectively. These packages would naturally have fewer potential users.
Counterfactual to this, the increase could be due to other external
factors, such as an increase in the number of skilled developers (and
thus more package contributions), a change in CRAN policy or management
to add packages (either to CRAN and/or task view), and/or the fact that
new packages are still yet to amass users. While there is an argument
that low download counts are due to the low utility and/or quality of
packages, packages in CRAN task views are selected by expert
maintainers. We can reasonably assume that any package listed in the
CRAN task view has an acceptable utility and quality.

If the downloads are reflective of the experimental designs used in
practice, \textbf{\emph{a small set of packages appears to be sufficient
for most users to construct the full set of designs of experiments they
need in practice}}. Packages of course evolve, and the top downloaded
packages have had regular updates that may have broadened their scope
from their previous releases.

\begin{figure}[htbp]

{\centering \includegraphics{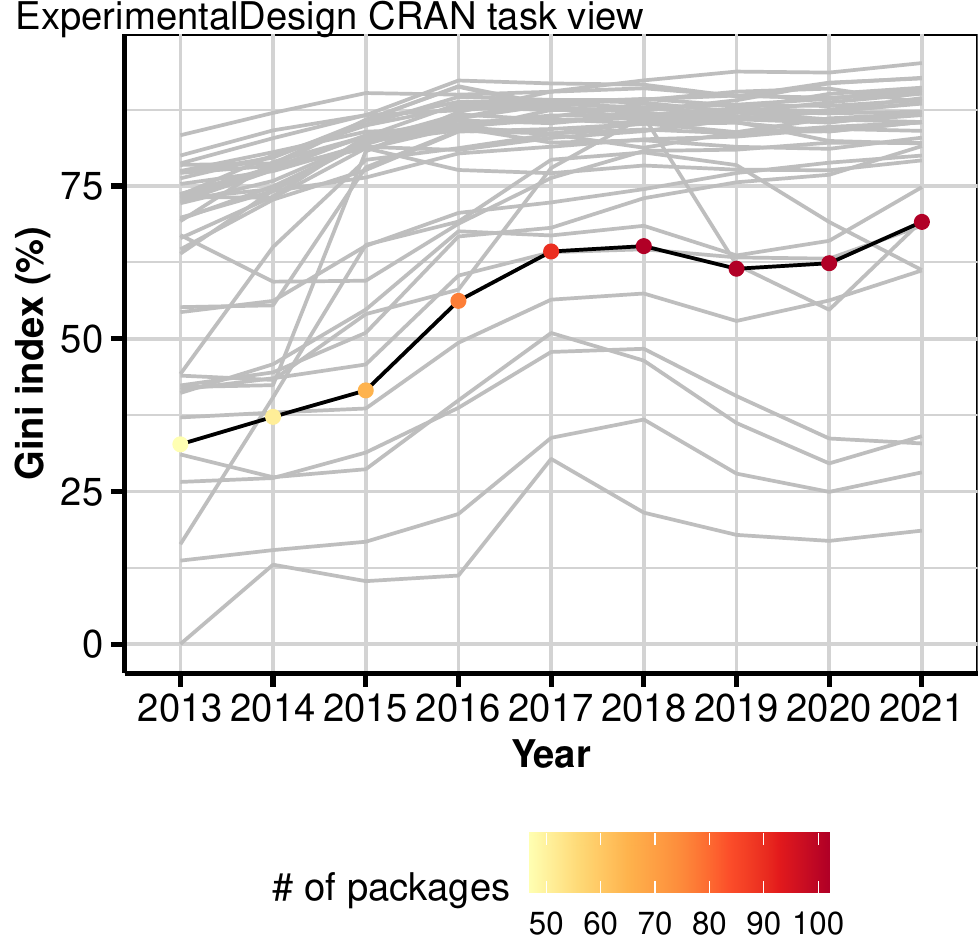} 

}

\caption{The points show the Gini index of the download counts by year for the ExperimentalDesign task view with the color showing the number of packages. The gray lines show the line plots of the Gini index across years for all other CRAN task views. See Figure S1 in the Supplementary Material for the line graph of the Gini index across years for each CRAN task view.}\label{fig:fig-gini-all-ctvs}
\end{figure}

While in absolute terms the Gini index is high for the
\href{http://CRAN.R-project.org/view=ExperimentalDesign}{\emph{ExperimentalDesign}}
task view (32.7\% to 69.1\%), the inequality is not as severe as in
other CRAN task views as shown in Figure \ref{fig:fig-gini-all-ctvs}. We
can see in Figure \ref{fig:fig-gini-all-ctvs} that the Gini index
generally increases over time for DoE packages (as is generally the case
for other CRAN task views as shown in Figure S1 in the Supplementary
Materials), but most other CRAN task views have a Gini index of over
75\%. This suggests that other CRAN task views may have dominant
standards, and in comparison to other topics, there are more
\textbf{\emph{diverse approaches to designing experiments}}, and thus,
no single DoE package is dominant. However, this observation doe not
consider other approaches to generate experimental designs, such as the
proprietary software, CycDesignN (Whittaker, Williams, and John 2022),
which may be widely used.

\hypertarget{the-field-of-experimental-design-is-slow-changing}{%
\subsection{The field of experimental design is
slow-changing}\label{the-field-of-experimental-design-is-slow-changing}}

\begin{figure}[htbp]

{\centering \includegraphics{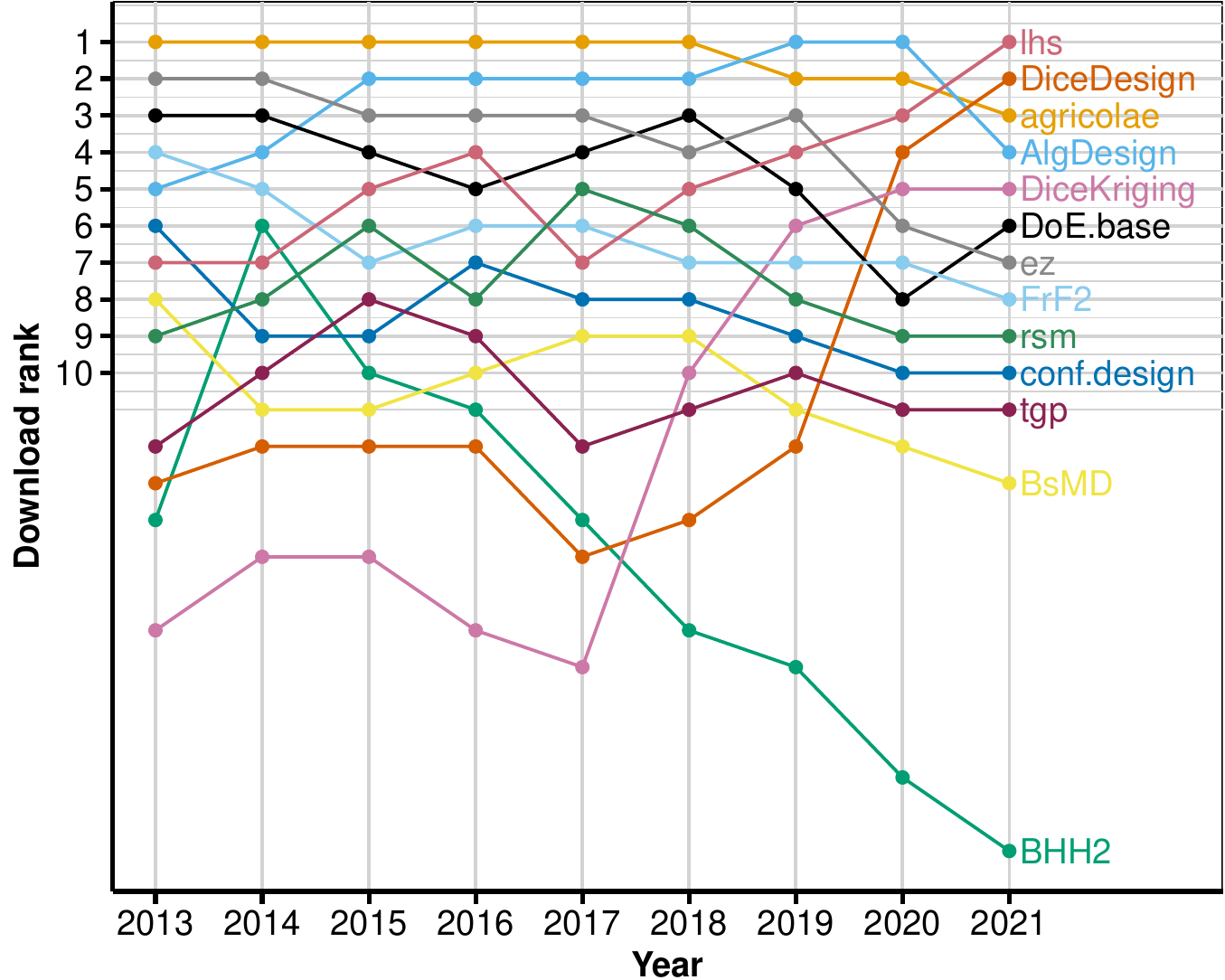} 

}

\caption{The plot shows the rank of the top 10 packages downloaded by year. Packages that did not appear in the top 10 for at least two periods were omitted from the plot. Most packages are consistently in the top 10 for the period shown.}\label{fig:rank-over-time}
\end{figure}

\begin{figure}[htbp]

{\centering \includegraphics{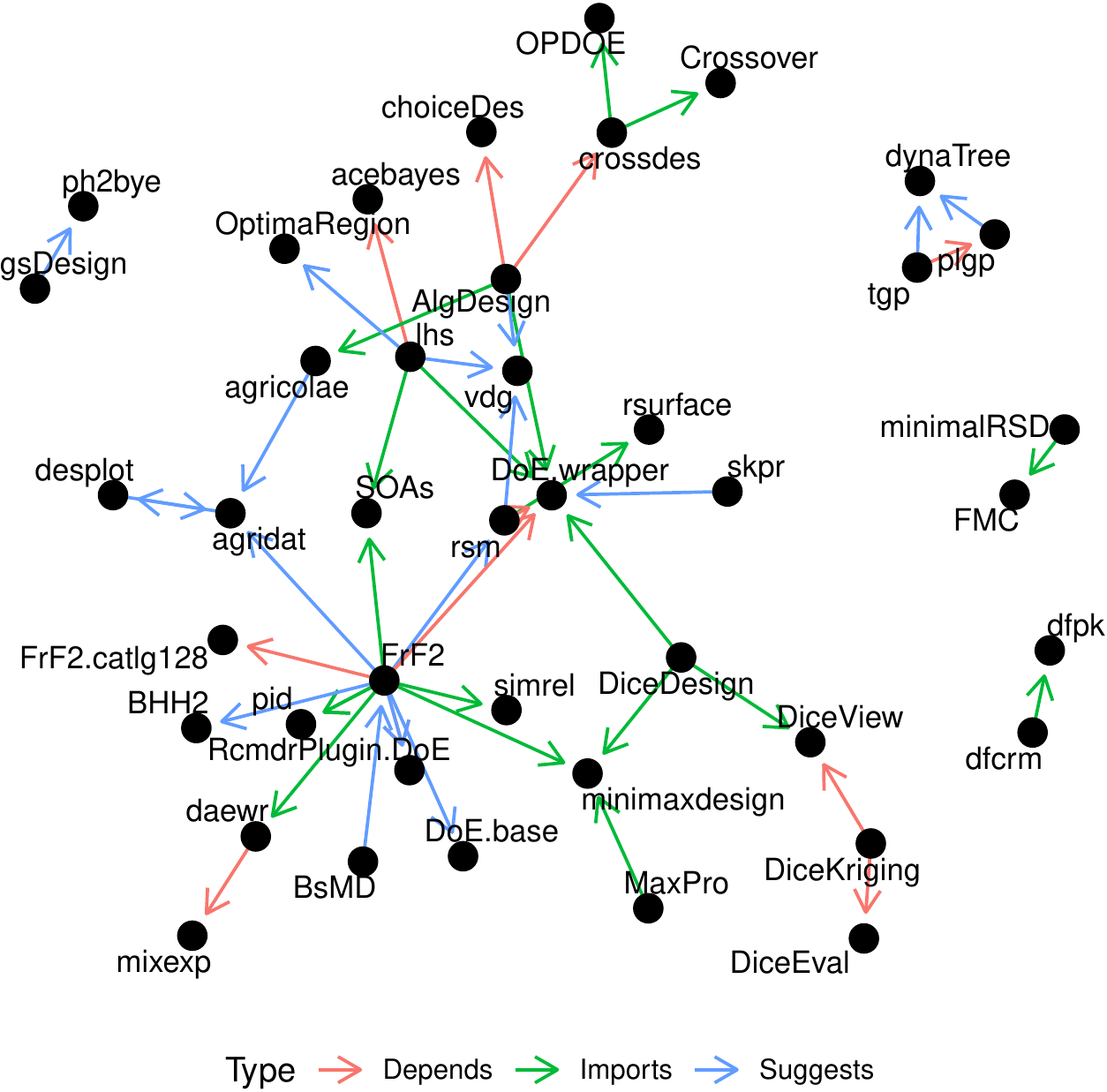} 

}

\caption{Package connections (depends, suggests and imports) within DoE packages. The direction of the arrow shows the connection of packags where the package on the tail of the arrow is a dependency, suggestion or import for the package on the head of the arrow.  DoE packages that do not depend, suggest or import another DoE package are not shown.}\label{fig:plot-doe-network}
\end{figure}

We can see in Figure \ref{fig:rank-over-time} that most of the top 10
ranking packages have been in the top 10 for the last nine years with
\href{https://cran.r-project.org/web/packages/lhs/index.html}{\texttt{lhs}}
steadily climbing up the ranks in the last few years. It should be noted
that the download of one package can prompt the download of another
package; the most notable package connection is
\href{https://cran.r-project.org/web/packages/AlgDesign/index.html}{\texttt{AlgDesign}}
and
\href{https://cran.r-project.org/web/packages/agricolae/index.html}{\texttt{agricolae}},
where the former is an import for the latter. The full network of
package connections within the DoE packages is shown in Figure
\ref{fig:plot-doe-network}.

\begin{figure}[htbp]

{\centering \includegraphics{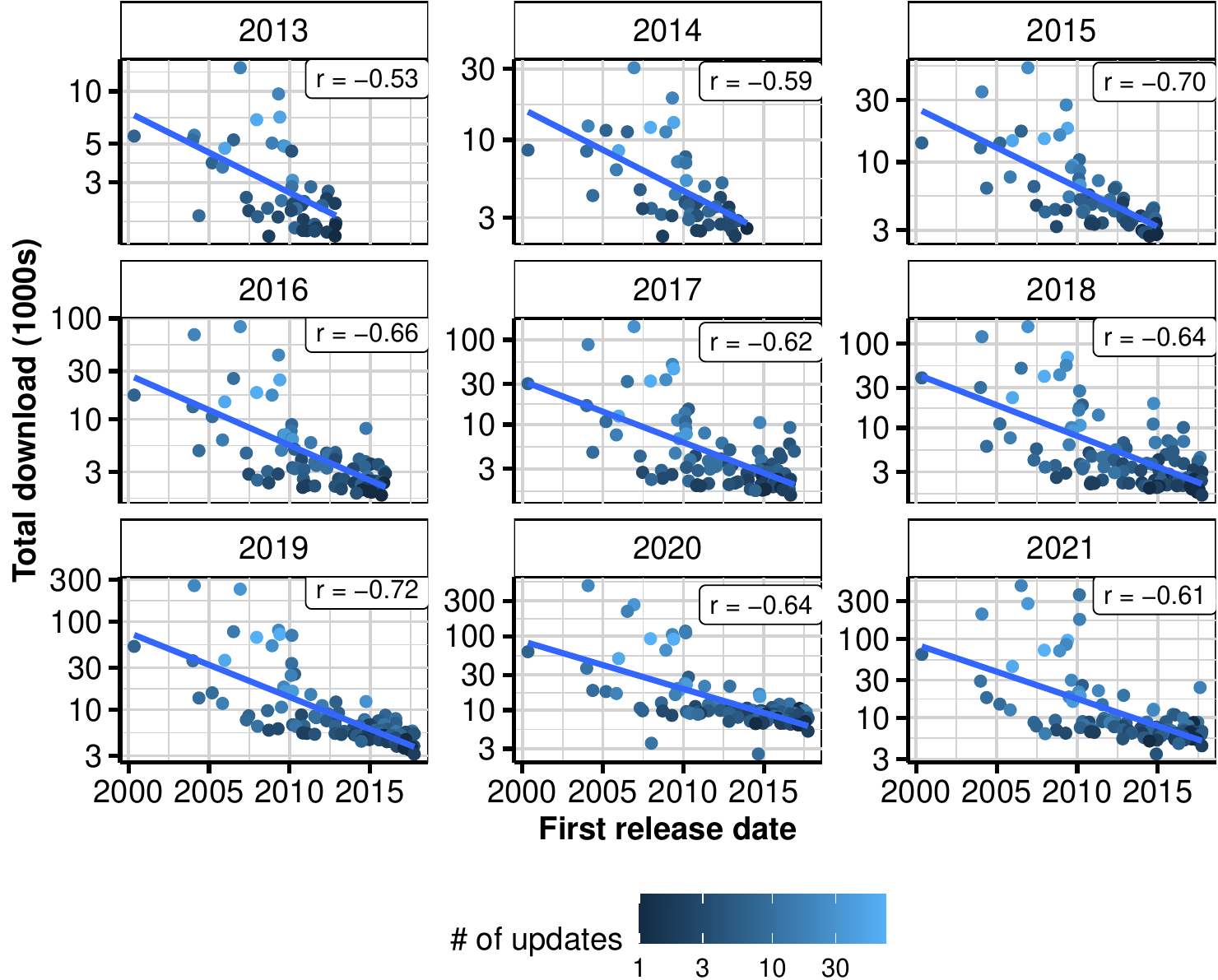} 

}

\caption{The above figure shows the total download (in log scale) of a package in the corresponding year against the first release date of the package. The blue line corresponds to the least-squares fit of the simple linear regression model. The label in the upper right-hand corner shows the sample correlation coefficient between the first release date and log (with base 10) of the total download count. The high leverage point on the far left belongs to `conf.design`, authored by one of the earlier contributors to R.}\label{fig:release-date-vs-download}
\end{figure}

Figure \ref{fig:release-date-vs-download} shows a moderate negative
correlation between the first release date and the (log of) total
download counts of DoE packages for any given year from 2013 to 2021.
This suggests that packages released earlier are more likely to be used
today (possibly for legacy reasons or the general inertia to adopt new
packages). We can also see in Figure \ref{fig:release-date-vs-download}
that most downloaded packages were released from 2004 to 2010.

The consistency in the top 10 ranking packages (Figure
\ref{fig:rank-over-time}) and the fact that most downloaded DoE packages
were first released more than 10 years ago (Figure
\ref{fig:release-date-vs-download}) indicate that either the existing
packages fulfilled the needs of mass in practice or no new packages were
compelling for many to switch their practice. However, we also see that
the top downloaded packages generally have more updates (see Figure
\ref{fig:release-date-vs-download}); therefore, it is possible that the
packages have improved or broadened the scope of their usage.

\hypertarget{optimal-designs-are-of-interest}{%
\subsection{Optimal designs are of
interest}\label{optimal-designs-are-of-interest}}

\begin{figure}[htbp]

{\centering \includegraphics{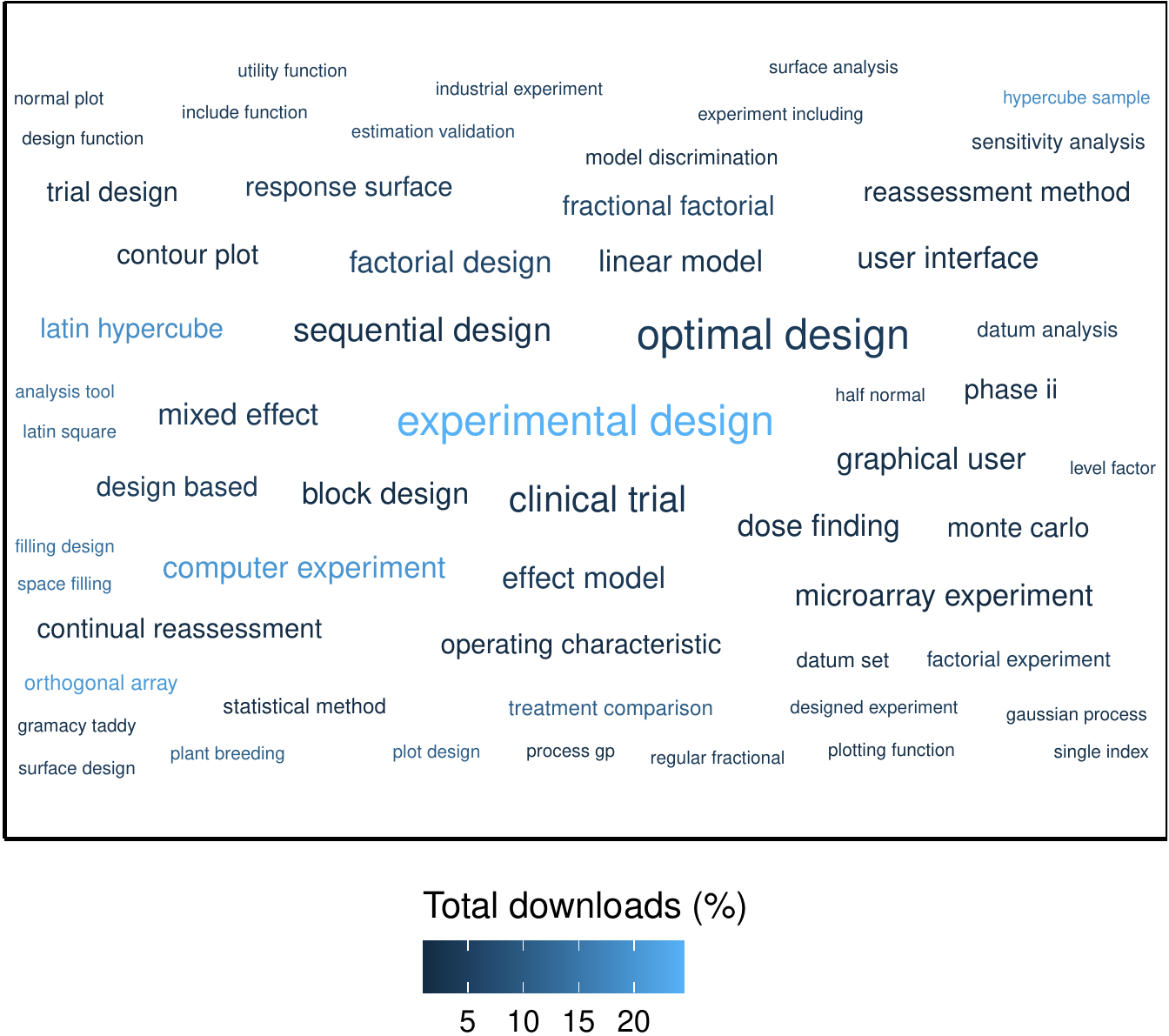} 

}

\caption{The above figure shows the word cloud of bigrams from the title and descriptions of the DoE packages. The size shows how often the bigram appears across the DoE packages and the color is relative to the total download count in 2021 for the packages that contain the bigram.}\label{fig:wordcloud-over-time}
\end{figure}

Figure \ref{fig:wordcloud-over-time} shows some of the common purposes
of DoE packages, based on bigrams in the package title and description.
We only show the bigrams as unigrams were not insightful, and there were
not many trigrams common across packages. To count the bigrams, we
processed the text data as follows:

\begin{enumerate}
\def\labelenumi{\arabic{enumi}.}
\tightlist
\item
  We standardized the words to the lower case and removed pluralization.
\item
  Multiple mentions of the same bigram within a package were counted as
  one (for example,
  \href{https://cran.r-project.org/web/packages/AlgDesign/index.html}{\texttt{AlgDesign}}
  mentions ``experimental design'' four times in the title and the
  description, but this is counted as one).
\item
  Bigrams consisting of stop words were removed. The stop words are
  sourced from the lexicons in \texttt{tidytext::stop\_words} in
  addition to other words we deemed irrelevant, e.g.~``provide'',
  ``e.g.'', ``calculate'' and so on -- full list is shown in the code
  provided in the link under Section \ref{pkgs}.
\end{enumerate}

Unsurprisingly, the bigram ``experimental design'' was the most common.
More interestingly, ``optimal design'' and ``sequential design''
appeared across different packages (indicated by the size of the word in
Figure \ref{fig:wordcloud-over-time}), and the bigrams ``latin
hypercube'' and ``computer experiment'' are used across a few packages
that are downloaded frequently (indicated by the color of the word in
Figure \ref{fig:wordcloud-over-time}). Sequential design, Latin
hypercube sampling and computer experiments (which generally include
space-filling designs such as Latin hypercube sampling) generally
operate by optimizing a user-selected criterion and can be classified as
optimal designs.

Although there exists a separate
\href{http://CRAN.R-project.org/view=ClinicalTrials}{\emph{ClinicalTrials}}
task view, the DoE packages clearly include some packages that are of
interest to clinical trials as shown by the size of the bigram
``clinical trial'' (and related bigrams like ``dose finding'' and
``phase ii'') in Figure \ref{fig:wordcloud-over-time}.

\hypertarget{development-of-experimental-designs-occur-in-silos}{%
\subsection{Development of experimental designs occur in
silos}\label{development-of-experimental-designs-occur-in-silos}}

\begin{figure}[htbp]

{\centering \includegraphics{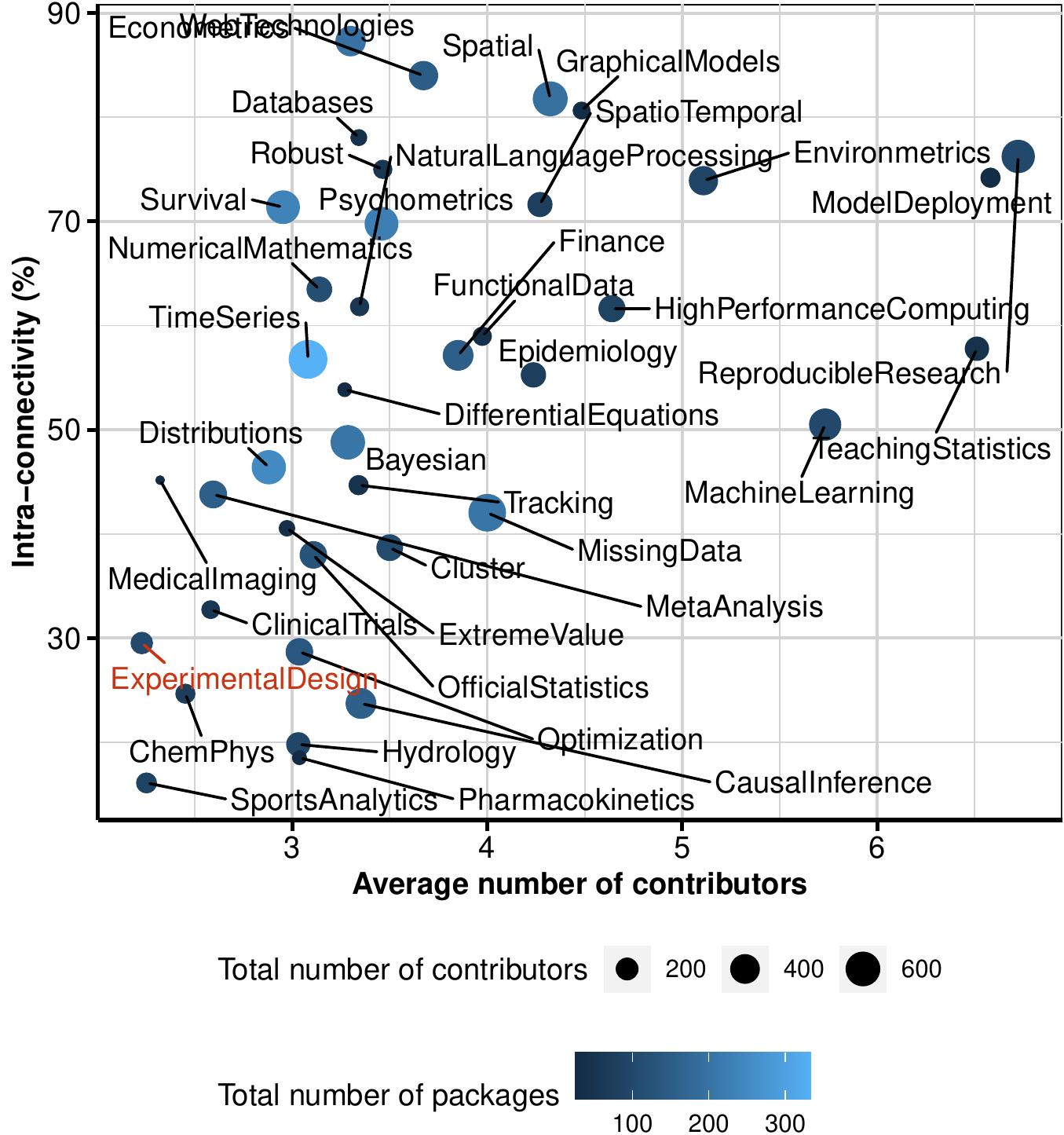} 

}

\caption{The above figure is a scatterplot of intra-connectivity (the percentage of packages that depends, suggests or imports at least one other package within the same task view) and the average number of contributors for each CRAN task view. Low intra-connectivity suggests that development within the topic mostly occurs in silos, while high  intra-connectivity suggests that there are more interactions within the topic. The color shows the number of packages, the size of the point corresponds to the total number of contributors, and the text labels show the CRAN task view names.  The label of ExperimentalDesign task view is colored in red. The task views in the bottom-left corner are topics that are more indicative of contributors working in silos. The actual numerical values are listed in Table S1 in the Supplementary Material.}\label{fig:ctv-summ-plot}
\end{figure}

Figure \ref{fig:ctv-summ-plot} shows that the
\href{http://CRAN.R-project.org/view=ExperimentalDesign}{\emph{ExperimentalDesign}}
task view has the lowest average number of contributors among all the 39
CRAN task views. In addition, we can also see in Figure
\ref{fig:ctv-summ-plot} that the
\href{http://CRAN.R-project.org/view=ExperimentalDesign}{\emph{ExperimentalDesign}}
task view has one of the least intra-connectivity (the percentage of
packages that make use of other packages within the same task view). The
full connection between DoE packages is shown in Figure
\ref{fig:plot-doe-network}. These observations suggest that
\textbf{\emph{experimental design is one of the least collaborative
fields}} and package development generally occurs in silos.

\hypertarget{design}{%
\section{Interface design}\label{design}}

In software design, there are two interface designs to consider: user
interface (UI) and application programming interface (API). The UI is
concerned with the interaction of the software by the user, while the
API is concerned with how different programs interact and is
predominately of interest to the developer. The UI design is an
abstraction that specifies the desired experimental design, and its
choices enable how a user expresses the specification of an experimental
design. The API design aids other developers in leveraging existing
systems.

In this section, we discuss the interface designs of functions that
output an experimental design based on three broad areas: factorial,
recipe and augmenting designs. The discussion is exclusive to the top
downloaded packages (shown in Figure \ref{fig:rank-over-time}), with the
exception of
\href{https://cran.r-project.org/web/packages/ez/index.html}{\texttt{ez}}
and
\href{https://cran.r-project.org/web/packages/DiceKriging/index.html}{\texttt{DiceKriging}},
as the former is predominately visualization of experimental data and
the latter is about the analysis of computer experiments in addition to
belonging to the same suite of packages as
\href{https://cran.r-project.org/web/packages/DiceDesign/index.html}{\texttt{DiceDesign}}
(Dupuy, Helbert, and Franco 2015).

\hypertarget{the-case-of-factorial-designs}{%
\subsection{The case of factorial
designs}\label{the-case-of-factorial-designs}}

Factorial experiments offer a challenge in allocating the treatment
factors to experimental units where the full set of factorial treatments
cannot be administered (and replicated) and/or the experimental units
have a grouping structure. The effort to address this challenge is
reflected in the number of packages that focus on the construction of
factorial designs as described next.

The
\href{https://cran.r-project.org/web/packages/DoE.base/index.html}{\texttt{DoE.base}}
package (Grömping 2018) can construct full factorial and (regular and
irregular) orthogonal array designs via \texttt{fac.design()} and
\texttt{oa.design()}, respectively. The
\href{https://cran.r-project.org/web/packages/FrF2/index.html}{\texttt{FrF2}}
package (Grömping 2014) constructs regular fractional 2-level factorial
designs using \texttt{FrF2()} and \texttt{FrF2Large()}, with the latter
for large designs. The
\href{https://cran.r-project.org/web/packages/conf.design/index.html}{\texttt{conf.design}}
package (Venables 2013) constructs symmetric confounded factorial
designs via \texttt{conf.design()} and the
\href{https://cran.r-project.org/web/packages/BHH2/index.html}{\texttt{BHH2}}
package (Barrios 2016) generates a full or fractional 2-level factorial
design matrix via \texttt{ffDesMatrix()}.

While the argument names and underlying algorithms of these functions to
generate factorial designs differ, it generally requires the users to
input the number of:

\begin{itemize}
\tightlist
\item
  treatment factors,
\item
  experimental runs (or replications), and
\item
  levels for each factor if the design was allowed to vary in the number
  of levels.
\end{itemize}

The output of the design is either a special class of list (e.g.~the
\texttt{design} class for
\href{https://cran.r-project.org/web/packages/DoE.base/index.html}{\texttt{DoE.base}}
and
\href{https://cran.r-project.org/web/packages/FrF2/index.html}{\texttt{FrF2}}
or a \texttt{data.frame} for
\href{https://cran.r-project.org/web/packages/conf.design/index.html}{\texttt{conf.design}})
or a matrix for \texttt{BHH2} such that an element or column corresponds
to a treatment factor with each value corresponding to one experimental
run. The treatment factors are generally assigned pseudo names
(e.g.~letters of the alphabet) or an argument exists for users to input
treatment names as a character vector.

Some forms of factorial design are known by other names, for example,
\emph{response surface designs} are factorial designs where the
treatment factors are discrete levels of continuous variables. Two types
of response surface designs can be constructed using the
\href{https://cran.r-project.org/web/packages/rsm/index.html}{\texttt{rsm}}
package (Lenth 2009): Box-Behnken design (Box and Behnken 1960) and
central-composite designs (Box and Wilson 1951) via functions
\texttt{bbd()} and \texttt{ccd()}, respectively, where the minimum
required input is the number of factors. Another form of factorial
design is the \emph{saturated designs} where higher-order interaction
effects of treatment factors are typically confounded with the main
effects. Plankett-Burnman designs (Plackett and Burman 1946) are a type
of saturated design that can be generated by the function \texttt{pb()}
in the
\href{https://cran.r-project.org/web/packages/FrF2/index.html}{\texttt{FrF2}}
package, where the user provides the number of experimental runs and the
number of treatment factors.

\hypertarget{the-case-of-recipe-designs}{%
\subsection{The case of recipe
designs}\label{the-case-of-recipe-designs}}

The
\href{https://cran.r-project.org/web/packages/agricolae/index.html}{\texttt{agricolae}}
package (de Mendiburu 2021) is the prime example of constructing designs
based on a set of so-called ``recipe functions'', where each function
corresponds to a single class of experimental design. For example,
\texttt{design.crd()}, \texttt{design.rcbd()} and
\texttt{design.split()} construct completely randomized, randomized
complete block and split-plot designs, respectively. Users typically
supply treatment labels (or the number of treatments in the case of
\texttt{design.split()}) and the number of replications as arguments for
these functions. The output is a list with one element corresponding to
a \texttt{data.frame} that contains the design in a table such that the
row corresponds to the experimental run and the columns correspond to
the experimental variables (we refer to this format simply as ``table
format'' henceforth).

The use of recipe functions is not limited to classical experimental
designs; the
\href{https://cran.r-project.org/web/packages/AlgDesign/index.html}{\texttt{AlgDesign}}
(Wheeler 2022) package offers three primary functions for generating
optimal designs: \texttt{optBlock()}, \texttt{optFederov()} and
\texttt{optMonteCarlo()}. In general, these functions require data and
formulas in terms of the supplied data variables with the choice of the
criterion (e.g.~the D-criterion), with the output as a list with one
element corresponding to the design in a table format. The difference
between these functions lies in the underlying search strategy for
optimal designs, and the name of the function is a surrogate for the
search algorithm.

Computer experiments, which generally involve space-filling designs, are
implemented in packages such as
\href{https://cran.r-project.org/web/packages/lhs/index.html}{\texttt{lhs}}
(Carnell 2022) and
\href{https://cran.r-project.org/web/packages/DiceDesign/index.html}{\texttt{DiceDesign}}
(Dupuy, Helbert, and Franco 2015). For the
\href{https://cran.r-project.org/web/packages/lhs/index.html}{\texttt{lhs}}
package, functions such as \texttt{randomLHS()}, \texttt{optimumLHS()},
and \texttt{maximinLHS()}, require users to specify the sample size
(\(n\)) and the number of variables (\(p\)), then it generates a Latin
hypercube sample (McKay, Beckman, and Conover 1979), based on different
optimization schemes (in this case, random, S optimal and maxmin
criteria, respectively; see package documentation for more details).
Similarly for
\href{https://cran.r-project.org/web/packages/DiceDesign/index.html}{\texttt{DiceDesign}},
there is a comprehensive list of space-filling designs such as
\texttt{dmaxDesign()}, \texttt{lhsDesign()}, and \texttt{wspDesign()}
with input of \(n\) and \(p\) as before (among additional parameters for
some) that implement algorithms that either maximize the entropy (Shewry
and Wynn 1987), produce a random Latin hypercube sample, or use the WSP
algorithm (Santiago, Claeys-Bruno, and Sergent 2012), respectively.
These designs output an \(n \times p\) matrix with values between 0 and
1. Again, these functions employ a recipe style, in which each function
has a name that corresponds to a certain search strategy to generate the
experimental design.

\hypertarget{the-case-of-augmenting-designs}{%
\subsection{The case of augmenting
designs}\label{the-case-of-augmenting-designs}}

Some functions in the DoE packages require the input of existing
experimental designs to produce new designs. For example, the
\href{https://cran.r-project.org/web/packages/DoE.base/index.html}{\texttt{DoE.base}}
package contains some experimental functions, \texttt{cross.design()}
and \texttt{param.design()}, to combine designs with the former taking a
Cartesian product of the input designs while the latter uses a Taguchi
style (Taguchi 1986) to aggregate the designs with the inner and outer
arrays. The
\href{https://cran.r-project.org/web/packages/lhs/index.html}{\texttt{lhs}}
package contains a function that \texttt{augmentLHS()} to add additional
samples to the existing Latin hypercube sample.

Another class of augmenting design is \emph{sequential design} (also
called \emph{adaptive sampling}), which is best represented by
\href{https://cran.r-project.org/web/packages/tgp/index.html}{\texttt{tgp}}
(Gramacy and Taddy 2010). This requires prior information that is used
to inform the next experimental design using \texttt{tgp.design()} and
\texttt{dopt.gp()}. The user is required to supply candidate samples to
subsample from and a model or a prior experimental design. Follow-up
experiments, which can also be classified as sequential designs, is
implemented by
\href{https://cran.r-project.org/web/packages/BsMD/index.html}{\texttt{BsMD}}
(Barrios 2020) using a model-discriminant approach using \texttt{MD()}.

\hypertarget{discussion}{%
\section{Discussion}\label{discussion}}

Through the exploratory data analysis of the three data sources (package
download logs, package metadata and CRAN task views) outlined in Section
\ref{data}, we observed in Section \ref{eda} that the the total download
of DoE packages is concentrated only on a handful of R-packages although
these represent a diverse set in comparison with other CRAN task views.
Furthermore, the data suggest that experimental design is the least
collaborative field.

There are a number of limitations and shortcomings to our exploratory
data analysis. First, CRAN task views are volunteer maintained so some
experimental design packages may not be included in the DoE packages.
Second, we used only the RStudio CRAN mirror download, which may have
biased our observations. Third, our analysis was limited to R-packages
alone and many practitioners may use other methods to construct
experimental designs. Finally, all our statements should be treated as
speculative rather than conclusive; the data are all observational so no
conclusive, generalizable statement is possible. Regardless, the
data-driven nature of our analysis provides objective insight into the
field of experimental design.

The interface design (discussed in Section \ref{design}) reveals that
the most widely used DoE packages generally have functions that 1) focus
on certain aspects of experimental design (e.g.~factorial structure or
augmenting design), 2) are a recipe format (i.e.~the name of the
function is a surrogate to a single class of the design or optimal
search algorithm) and 3) context is often a second thought -- many
inputs a single integer corresponding to the number of factors, levels,
or experimental runs (or sample size). The function will often assign
pseudo-factor names or there is an optional argument to input a
character vector that corresponds to the factor names. These interface
designs require users to have processed the experiment in statistical
terms (often stripping the experimental context away) and
simultaneously, users must choose the generation mechanism by selecting
an appropriate function. Arguably, the current dominant interface
designs are not aligned with the way practitioners cognitively design
their experiments. Often, the critical part of designing an experiment
is to understand the experimental structure, and the experimental
context can govern or guide the choice of algorithm to allocate
treatments. In addition, the nature of recipe functions can obscure the
understanding and relation of designs (e.g., how do you go from an
unstructured factorial design to split plot design?). Each new method
for generating an experimental design appears to correspond to a
completely new function, and intermediate results are often not easily
accessible. These factors may contribute to why developers often work in
silos in the field of experimental design.

A consistent cognitive interface design that leverages existing
developments and can be easily extended to new methods will conceivably
be of great help to practitioners. Some efforts to this end are seen in
\href{https://cran.r-project.org/web/packages/DoE.wrapper/index.html}{\texttt{DoE.wrapper}}
(Grömping 2020), which contains wrapper recipe functions for other DoE
packages such as
\href{https://cran.r-project.org/web/packages/lhs/index.html}{\texttt{lhs}},
\href{https://cran.r-project.org/web/packages/AlgDesign/index.html}{\texttt{AlgDesign}}
and
\href{https://cran.r-project.org/web/packages/FrF2/index.html}{\texttt{FrF2}}
(see Figure \ref{fig:plot-doe-network}), and is also the subject of the
developmental package
\href{https://cran.r-project.org/web/packages/edibble/index.html}{\texttt{edibble}}
(Tanaka 2021). Undoubtedly, no single developer or package can cater to
all experimental designs; therefore, any unifying interface should
consider how other developers can contribute or add their methods.
Future research could benefit from further exploratory data analysis,
expanding the study to beyond R-packages, and discussing other aspects
of interface designs.

\hypertarget{pkgs}{%
\section{Acknowledgement}\label{pkgs}}

This paper uses
\href{https://cran.r-project.org/web/packages/targets/index.html}{\texttt{targets}}
(Landau 2021) and
\href{https://cran.r-project.org/web/packages/renv/index.html}{\texttt{renv}}
(Ushey 2022) for reproducibility,
\href{https://cran.r-project.org/web/packages/knitr/index.html}{\texttt{knitr}}
(Xie 2015) and
\href{https://cran.r-project.org/web/packages/rmarkdown/index.html}{\texttt{rmarkdown}}
(Xie, Allaire, and Grolemund 2018) for creating reproducible documents,
\href{https://cran.r-project.org/web/packages/ggplot2/index.html}{\texttt{ggplot2}}
(Wickham 2016),
\href{https://cran.r-project.org/web/packages/ggraph/index.html}{\texttt{ggraph}}
(Pedersen 2021),
\href{https://cran.r-project.org/web/packages/ggwordcloud/index.html}{\texttt{ggwordcloud}}
(Le Pennec and Slowikowski 2019) and \texttt{colorspace} (Zeileis et al.
2020) for visualization,
\href{https://cran.r-project.org/web/packages/kableExtra/index.html}{\texttt{kableExtra}}
(Zhu 2021) for customizing the table in the Supplementary Material and
\href{https://cran.r-project.org/web/packages/tidyverse/index.html}{\texttt{tidyverse}}
(Wickham et al. 2019),
\href{https://cran.r-project.org/web/packages/tidytext/index.html}{\texttt{tidytext}}
(Silge and Robinson 2016),
\href{https://cran.r-project.org/web/packages/pluralize/index.html}{\texttt{pluralize}}
(Rudis and Embrey 2020), and
\href{https://cran.r-project.org/web/packages/ineq/index.html}{\texttt{ineq}}
(Zeileis 2014) for data processing and manipulation, and
\href{https://cran.r-project.org/web/packages/cranlogs/index.html}{\texttt{cranlogs}}
(Csárdi 2019) and
\href{https://cran.r-project.org/web/packages/ctv/index.html}{\texttt{ctv}}
(Zeileis 2005) for extracting data. All codes used to reproduce this
paper are found in \url{https://github.com/emitanaka/paper-DoE-review}.

\hypertarget{refs}{}
\begin{CSLReferences}{1}{0}
\leavevmode\vadjust pre{\hypertarget{ref-BHH2}{}}%
Barrios, Ernesto. 2016. \emph{BHH2: Useful Functions for Box, Hunter and
Hunter II}. \url{https://CRAN.R-project.org/package=BHH2}.

\leavevmode\vadjust pre{\hypertarget{ref-BsMD}{}}%
---------. 2020. \emph{BsMD: Bayes Screening and Model Discrimination}.
\url{https://CRAN.R-project.org/package=BsMD}.

\leavevmode\vadjust pre{\hypertarget{ref-Box1960-wr}{}}%
Box, G E P, and D W Behnken. 1960. {``Some New Three Level Designs for
the Study of Quantitative Variables.''} \emph{Technometrics: A Journal
of Statistics for the Physical, Chemical, and Engineering Sciences} 2
(4): 455--75. \url{https://doi.org/10.2307/1266454}.

\leavevmode\vadjust pre{\hypertarget{ref-Box1951-ji}{}}%
Box, G E P, and K B Wilson. 1951. {``On the Experimental Attainment of
Optimum Conditions.''} \emph{Journal of the Royal Statistical Society.
Series B, Statistical Methodology} 13 (1): 1--45.

\leavevmode\vadjust pre{\hypertarget{ref-lhs}{}}%
Carnell, Rob. 2022. \emph{Lhs: Latin Hypercube Samples}.
\url{https://CRAN.R-project.org/package=lhs}.

\leavevmode\vadjust pre{\hypertarget{ref-cranlogs}{}}%
Csárdi, Gábor. 2019. \emph{Cranlogs: Download Logs from the 'RStudio'
'CRAN' Mirror}. \url{https://CRAN.R-project.org/package=cranlogs}.

\leavevmode\vadjust pre{\hypertarget{ref-agricolae}{}}%
de Mendiburu, Felipe. 2021. \emph{Agricolae: Statistical Procedures for
Agricultural Research}.
\url{https://CRAN.R-project.org/package=agricolae}.

\leavevmode\vadjust pre{\hypertarget{ref-DiceDesign}{}}%
Dupuy, Delphine, Céline Helbert, and Jessica Franco. 2015.
{``{DiceDesign} and {DiceEval}: Two {R} Packages for Design and Analysis
of Computer Experiments.''} \emph{Journal of Statistical Software} 65
(11): 1--38. \url{https://www.jstatsoft.org/v65/i11/}.

\leavevmode\vadjust pre{\hypertarget{ref-Fisher1935-qc}{}}%
Fisher, Ronald. 1935. \emph{The Design of Experiments}. Oliver; Boyd.

\leavevmode\vadjust pre{\hypertarget{ref-Gini1921-mf}{}}%
Gini, Corrado. 1921. {``Measurement of Inequality of Incomes.''}
\emph{The Economic Journal} 31 (121): 124--26.
\url{https://doi.org/10.2307/2223319}.

\leavevmode\vadjust pre{\hypertarget{ref-tgp}{}}%
Gramacy, Robert B., and Matthew Taddy. 2010. {``Categorical Inputs,
Sensitivity Analysis, Optimization and Importance Tempering with {tgp}
Version 2, an {R} Package for Treed Gaussian Process Models.''}
\emph{Journal of Statistical Software} 33 (6): 1--48.
\url{https://doi.org/10.18637/jss.v033.i06}.

\leavevmode\vadjust pre{\hypertarget{ref-FrF2}{}}%
Grömping, Ulrike. 2014. {``{R} Package {FrF2} for Creating and Analyzing
Fractional Factorial 2-Level Designs.''} \emph{Journal of Statistical
Software} 56 (1): 1--56. \url{https://www.jstatsoft.org/v56/i01/}.

\leavevmode\vadjust pre{\hypertarget{ref-DoE.base}{}}%
---------. 2018. {``{R} Package {DoE.base} for Factorial Experiments.''}
\emph{Journal of Statistical Software} 85 (5): 1--41.
\url{https://doi.org/10.18637/jss.v085.i05}.

\leavevmode\vadjust pre{\hypertarget{ref-DoE.wrapper}{}}%
---------. 2020. \emph{DoE.wrapper: Wrapper Package for Design of
Experiments Functionality}.
\url{https://CRAN.R-project.org/package=DoE.wrapper}.

\leavevmode\vadjust pre{\hypertarget{ref-targets}{}}%
Landau, William Michael. 2021. {``The Targets r Package: A Dynamic
Make-Like Function-Oriented Pipeline Toolkit for Reproducibility and
High-Performance Computing.''} \emph{Journal of Open Source Software} 6
(57): 2959. \url{https://doi.org/10.21105/joss.02959}.

\leavevmode\vadjust pre{\hypertarget{ref-ggwordcloud}{}}%
Le Pennec, Erwan, and Kamil Slowikowski. 2019. \emph{Ggwordcloud: A Word
Cloud Geom for 'Ggplot2'}.
\url{https://CRAN.R-project.org/package=ggwordcloud}.

\leavevmode\vadjust pre{\hypertarget{ref-rsm}{}}%
Lenth, Russell V. 2009. {``Response-Surface Methods in {R}, Using
{rsm}.''} \emph{Journal of Statistical Software} 32 (7): 1--17.
\url{https://doi.org/10.18637/jss.v032.i07}.

\leavevmode\vadjust pre{\hypertarget{ref-Lorenz1905-tc}{}}%
Lorenz, M O. 1905. {``Methods of Measuring the Concentration of
Wealth.''} \emph{Publications of the American Statistical Association} 9
(70): 209--19. \url{https://doi.org/10.2307/2276207}.

\leavevmode\vadjust pre{\hypertarget{ref-McKay1979-aw}{}}%
McKay, M D, R J Beckman, and W J Conover. 1979. {``Comparison of Three
Methods for Selecting Values of Input Variables in the Analysis of
Output from a Computer Code.''} \emph{Technometrics: A Journal of
Statistics for the Physical, Chemical, and Engineering Sciences} 21 (2):
239--45. \url{https://doi.org/10.1080/00401706.1979.10489755}.

\leavevmode\vadjust pre{\hypertarget{ref-ggraph}{}}%
Pedersen, Thomas Lin. 2021. \emph{Ggraph: An Implementation of Grammar
of Graphics for Graphs and Networks}.
\url{https://CRAN.R-project.org/package=ggraph}.

\leavevmode\vadjust pre{\hypertarget{ref-Plackett1946-ly}{}}%
Plackett, R L, and J P Burman. 1946. {``The Design of Optimum
Multifactorial Experiments.''} \emph{Biometrika} 33 (4): 302--25.

\leavevmode\vadjust pre{\hypertarget{ref-Pukelsheim2006-sv}{}}%
Pukelsheim, Friedrich. 2006. \emph{Optimal Design of Experiments}.

\leavevmode\vadjust pre{\hypertarget{ref-R}{}}%
R Core Team. 2021. \emph{R: A Language and Environment for Statistical
Computing}. Vienna, Austria: R Foundation for Statistical Computing.
\url{https://www.R-project.org/}.

\leavevmode\vadjust pre{\hypertarget{ref-python}{}}%
Rossum, G. van. 1995. {``Python Tutorial.''} CS-R9526. Amsterdam:
Centrum voor Wiskunde en Informatica (CWI).

\leavevmode\vadjust pre{\hypertarget{ref-pluralize}{}}%
Rudis, Bob, and Blake Embrey. 2020. \emph{Pluralize: Pluralize and
'Singularize' Any (English) Word}.
\url{https://CRAN.R-project.org/package=pluralize}.

\leavevmode\vadjust pre{\hypertarget{ref-Santiago2012-ui}{}}%
Santiago, J, M Claeys-Bruno, and M Sergent. 2012. {``Construction of
Space-Filling Designs Using {WSP} Algorithm for High Dimensional
Spaces.''} \emph{Chemometrics and Intelligent Laboratory Systems} 113
(April): 26--31. \url{https://doi.org/10.1016/j.chemolab.2011.06.003}.

\leavevmode\vadjust pre{\hypertarget{ref-sas1985sas}{}}%
SAS Institute. 1985. \emph{SAS User's Guide: Statistics}. Vol. 2. Sas
Inst.

\leavevmode\vadjust pre{\hypertarget{ref-Shewry1987-jn}{}}%
Shewry, M C, and H P Wynn. 1987. {``Maximum Entropy Sampling.''}
\emph{Journal of Applied Statistics} 14 (2): 165--70.
\url{https://doi.org/10.1080/02664768700000020}.

\leavevmode\vadjust pre{\hypertarget{ref-tidytext}{}}%
Silge, Julia, and David Robinson. 2016. {``Tidytext: Text Mining and
Analysis Using Tidy Data Principles in r.''} \emph{JOSS} 1 (3).
\url{https://doi.org/10.21105/joss.00037}.

\leavevmode\vadjust pre{\hypertarget{ref-Taguchi1986-ut}{}}%
Taguchi, Genichi. 1986. \emph{Introduction to Quality Engineering:
Designing Quality into Products and Processes}. Quality Resources.

\leavevmode\vadjust pre{\hypertarget{ref-edibble}{}}%
Tanaka, Emi. 2021. {``{edibble R-package}.''} \emph{GitHub Repository}.
\url{https://github.com/emitanaka/edibble}; GitHub.

\leavevmode\vadjust pre{\hypertarget{ref-renv}{}}%
Ushey, Kevin. 2022. \emph{Renv: Project Environments}.
\url{https://CRAN.R-project.org/package=renv}.

\leavevmode\vadjust pre{\hypertarget{ref-conf.design}{}}%
Venables, Bill. 2013. \emph{Conf.design: Construction of Factorial
Designs}. \url{https://CRAN.R-project.org/package=conf.design}.

\leavevmode\vadjust pre{\hypertarget{ref-Wasserman1975-xr}{}}%
Wasserman, Anthony I. 1975. {``Issues in Programming Language Design -
an Overview.''} In.

\leavevmode\vadjust pre{\hypertarget{ref-AlgDesign}{}}%
Wheeler, Bob. 2022. \emph{AlgDesign: Algorithmic Experimental Design}.
\url{https://CRAN.R-project.org/package=AlgDesign}.

\leavevmode\vadjust pre{\hypertarget{ref-cycdesignn}{}}%
Whittaker, D., E. R. Williams, and J. A John. 2022. \emph{CycDesign: A
Package for the Computer Generation of Experimental Designs,} (version
7.0). \url{https://vsni.co.uk/software/cycdesign}.

\leavevmode\vadjust pre{\hypertarget{ref-ggplot2}{}}%
Wickham, Hadley. 2016. \emph{Ggplot2: Elegant Graphics for Data
Analysis}. Springer-Verlag New York.
\url{https://ggplot2.tidyverse.org}.

\leavevmode\vadjust pre{\hypertarget{ref-Wickham2019-mj}{}}%
Wickham, Hadley, Mara Averick, Jennifer Bryan, Winston Chang, Lucy
D'agostino McGowan, Romain Francois, Garrett Grolemund, et al. 2019.
{``Welcome to the Tidyverse.''} \emph{Journal of Open Source Software} 4
(43): 1686.

\leavevmode\vadjust pre{\hypertarget{ref-knitr}{}}%
Xie, Yihui. 2015. \emph{Dynamic Documents with {R} and Knitr}. 2nd ed.
Boca Raton, Florida: Chapman; Hall/CRC. \url{https://yihui.org/knitr/}.

\leavevmode\vadjust pre{\hypertarget{ref-rmarkdown}{}}%
Xie, Yihui, J. J. Allaire, and Garrett Grolemund. 2018. \emph{R
Markdown: The Definitive Guide}. Boca Raton, Florida: Chapman; Hall/CRC.
\url{https://bookdown.org/yihui/rmarkdown}.

\leavevmode\vadjust pre{\hypertarget{ref-ctv}{}}%
Zeileis, Achim. 2005. {``{CRAN} Task Views.''} \emph{R News} 5 (1):
39--40. \url{https://CRAN.R-project.org/doc/Rnews/}.

\leavevmode\vadjust pre{\hypertarget{ref-ineq}{}}%
---------. 2014. \emph{Ineq: Measuring Inequality, Concentration, and
Poverty}. \url{https://CRAN.R-project.org/package=ineq}.

\leavevmode\vadjust pre{\hypertarget{ref-colorspace}{}}%
Zeileis, Achim, Jason C. Fisher, Kurt Hornik, Ross Ihaka, Claire D.
McWhite, Paul Murrell, Reto Stauffer, and Claus O. Wilke. 2020.
{``{colorspace}: A Toolbox for Manipulating and Assessing Colors and
Palettes.''} \emph{Journal of Statistical Software} 96 (1): 1--49.
\url{https://doi.org/10.18637/jss.v096.i01}.

\leavevmode\vadjust pre{\hypertarget{ref-kableExtra}{}}%
Zhu, Hao. 2021. \emph{kableExtra: Construct Complex Table with 'Kable'
and Pipe Syntax}. \url{https://CRAN.R-project.org/package=kableExtra}.

\end{CSLReferences}

\bibliographystyle{unsrt}
\bibliography{paper.bib}

@Manual{R,
    title = {R: A Language and Environment for Statistical Computing},
    author = {{R Core Team}},
    organization = {R Foundation for Statistical Computing},
    address = {Vienna, Austria},
    year = {2021},
    url = {https://www.R-project.org/},
  }

@Manual{renv,
    title = {renv: Project Environments},
    author = {Kevin Ushey},
    year = {2022},
    note = {R package version 0.15.5},
    url = {https://CRAN.R-project.org/package=renv},
  }

@Manual{ineq,
    title = {ineq: Measuring Inequality, Concentration, and Poverty},
    author = {Achim Zeileis},
    year = {2014},
    note = {R package version 0.2-13},
    url = {https://CRAN.R-project.org/package=ineq},
  }

@Manual{pluralize,
    title = {pluralize: Pluralize and 'Singularize' Any (English) Word},
    author = {Bob Rudis and Blake Embrey},
    year = {2020},
    note = {R package version 0.2.0},
    url = {https://CRAN.R-project.org/package=pluralize},
  }

@Manual{ggwordcloud,
    title = {ggwordcloud: A Word Cloud Geom for 'ggplot2'},
    author = {Erwan {Le Pennec} and Kamil Slowikowski},
    year = {2019},
    note = {R package version 0.5.0},
    url = {https://CRAN.R-project.org/package=ggwordcloud},
  }

@Article{tidytext,
    title = {tidytext: Text Mining and Analysis Using Tidy Data Principles in R},
    author = {Julia Silge and David Robinson},
    doi = {10.21105/joss.00037},
    url = {http://dx.doi.org/10.21105/joss.00037},
    year = {2016},
    publisher = {The Open Journal},
    volume = {1},
    number = {3},
    journal = {JOSS},
  }

@Article{colorspace,
    title = {{colorspace}: A Toolbox for Manipulating and Assessing Colors and Palettes},
    author = {Achim Zeileis and Jason C. Fisher and Kurt Hornik and Ross Ihaka and Claire D. McWhite and Paul Murrell and Reto Stauffer and Claus O. Wilke},
    journal = {Journal of Statistical Software},
    year = {2020},
    volume = {96},
    number = {1},
    pages = {1--49},
    doi = {10.18637/jss.v096.i01},
  }

@Manual{ggraph,
    title = {ggraph: An Implementation of Grammar of Graphics for Graphs and Networks},
    author = {Thomas Lin Pedersen},
    year = {2021},
    note = {R package version 2.0.5},
    url = {https://CRAN.R-project.org/package=ggraph},
  }

@Article{ctv,
    title = {{CRAN} Task Views},
    author = {Achim Zeileis},
    journal = {R News},
    year = {2005},
    volume = {5},
    number = {1},
    pages = {39--40},
    url = {https://CRAN.R-project.org/doc/Rnews/},
  }

@misc{edibble,
  author = {Tanaka, Emi},
  title = {{edibble R-package}},
  year = {2021},
  publisher = {GitHub},
  journal = {GitHub repository},
  howpublished = {\url{https://github.com/emitanaka/edibble}}
}

@Manual{cranlogs,
    title = {cranlogs: Download Logs from the 'RStudio' 'CRAN' Mirror},
    author = {Gábor Csárdi},
    year = {2019},
    note = {R package version 2.1.1},
    url = {https://CRAN.R-project.org/package=cranlogs},
}

@Article{targets,
    title = {The targets R package: a dynamic Make-like function-oriented pipeline toolkit for reproducibility and high-performance computing},
    author = {William Michael Landau},
    journal = {Journal of Open Source Software},
    year = {2021},
    volume = {6},
    number = {57},
    pages = {2959},
    url = {https://doi.org/10.21105/joss.02959},
  }

@Book{knitr,
    title = {Dynamic Documents with {R} and knitr},
    author = {Yihui Xie},
    publisher = {Chapman and Hall/CRC},
    address = {Boca Raton, Florida},
    year = {2015},
    edition = {2nd},
    note = {ISBN 978-1498716963},
    url = {https://yihui.org/knitr/},
  }

@Book{rmarkdown,
    title = {R Markdown: The Definitive Guide},
    author = {Yihui Xie and J.J. Allaire and Garrett Grolemund},
    publisher = {Chapman and Hall/CRC},
    address = {Boca Raton, Florida},
    year = {2018},
    note = {ISBN 9781138359338},
    url = {https://bookdown.org/yihui/rmarkdown},
  }

@Book{ggplot2,
    author = {Hadley Wickham},
    title = {ggplot2: Elegant Graphics for Data Analysis},
    publisher = {Springer-Verlag New York},
    year = {2016},
    isbn = {978-3-319-24277-4},
    url = {https://ggplot2.tidyverse.org},
  }

@Manual{kableExtra,
    title = {kableExtra: Construct Complex Table with 'kable' and Pipe Syntax},
    author = {Hao Zhu},
    year = {2021},
    note = {R package version 1.3.4},
    url = {https://CRAN.R-project.org/package=kableExtra},
  }

@Manual{lhs,
    title = {lhs: Latin Hypercube Samples},
    author = {Rob Carnell},
    year = {2022},
    note = {R package version 1.1.5},
    url = {https://CRAN.R-project.org/package=lhs},
}

@Manual{agricolae,
    title = {agricolae: Statistical Procedures for Agricultural Research},
    author = {Felipe {de Mendiburu}},
    year = {2021},
    note = {R package version 1.3-5},
    url = {https://CRAN.R-project.org/package=agricolae},
  }

@Manual{AlgDesign,
    title = {AlgDesign: Algorithmic Experimental Design},
    author = {Bob Wheeler},
    year = {2022},
    note = {R package version 1.2.1},
    url = {https://CRAN.R-project.org/package=AlgDesign},
  }

@Article{rsm,
    title = {Response-Surface Methods in {R}, Using {rsm}},
    author = {Russell V. Lenth},
    journal = {Journal of Statistical Software},
    year = {2009},
    volume = {32},
    number = {7},
    pages = {1--17},
    doi = {10.18637/jss.v032.i07},
  }

@Manual{conf.design,
    title = {conf.design: Construction of factorial designs},
    author = {Bill Venables},
    year = {2013},
    note = {R package version 2.0.0},
    url = {https://CRAN.R-project.org/package=conf.design},
  }

@Article{FrF2,
    title = {{R} Package {FrF2} for Creating and Analyzing Fractional Factorial 2-Level Designs},
    author = {Ulrike Gr{\"o}mping},
    journal = {Journal of Statistical Software},
    year = {2014},
    volume = {56},
    number = {1},
    pages = {1--56},
    url = {https://www.jstatsoft.org/v56/i01/},
  }

@Article{DiceDesign,
    title = {{DiceDesign} and {DiceEval}: Two {R} Packages for Design and Analysis of Computer Experiments},
    author = {Delphine Dupuy and C{\'e}line Helbert and Jessica Franco},
    journal = {Journal of Statistical Software},
    year = {2015},
    volume = {65},
    number = {11},
    pages = {1--38},
    url = {https://www.jstatsoft.org/v65/i11/},
  }

@Article{tgp,
    title = {Categorical Inputs, Sensitivity Analysis, Optimization and Importance Tempering with {tgp} Version 2, an {R} Package for Treed Gaussian Process Models},
    author = {Robert B. Gramacy and Matthew Taddy},
    journal = {Journal of Statistical Software},
    year = {2010},
    volume = {33},
    number = {6},
    pages = {1--48},
    url = {https://www.jstatsoft.org/v33/i06/},
    doi = {10.18637/jss.v033.i06},
  }

@Article{DiceKriging,
    title = {{DiceKriging}, {DiceOptim}: Two {R} Packages for the Analysis of Computer Experiments by Kriging-Based Metamodeling and Optimization},
    author = {Olivier Roustant and David Ginsbourger and Yves Deville},
    journal = {Journal of Statistical Software},
    year = {2012},
    volume = {51},
    number = {1},
    pages = {1--55},
    url = {https://www.jstatsoft.org/v51/i01/},
  }

@Article{DoE.base,
    title = {{R} Package {DoE.base} for Factorial Experiments},
    author = {Ulrike Gr{\"o}mping},
    journal = {Journal of Statistical Software},
    year = {2018},
    volume = {85},
    number = {5},
    pages = {1--41},
    doi = {10.18637/jss.v085.i05},
  }

@Manual{BsMD,
    title = {BsMD: Bayes Screening and Model Discrimination},
    author = {Ernesto Barrios},
    year = {2020},
    note = {R package version 2020.4.30},
    url = {https://CRAN.R-project.org/package=BsMD},
  }

@Manual{BHH2,
    title = {BHH2: Useful Functions for Box, Hunter and Hunter II},
    author = {Ernesto Barrios},
    year = {2016},
    note = {R package version 2016.05.31},
    url = {https://CRAN.R-project.org/package=BHH2},
  }

@Manual{DoE.wrapper,
    title = {DoE.wrapper: Wrapper Package for Design of Experiments Functionality},
    author = {Ulrike Gr{\"o}mping},
    year = {2020},
    note = {R package version 0.11},
    url = {https://CRAN.R-project.org/package=DoE.wrapper},
  }

\end{document}